\documentstyle[prl,aps]{revtex}

\newtheorem{theorem}{Theorem}

\newtheorem{definition}[theorem]{Definition}

\newtheorem{proposition}[theorem]{Proposition}

\begin{document}
\draft

\twocolumn[\hsize\textwidth\columnwidth\hsize\csname@twocolumnfalse\endcsname

\title{{\bf A stochastic limit approach to the SAT problem}
}

\author{Luigi  Accardi$^\dagger$ and Masanori Ohya$^\ddagger $
}

\address{
$\dagger$
Centro V.Volterra, Universit\`a di Roma Torvergata, 
Via Orazio Raimondo, 00173 Roma, Italia \\
e-mail: accardi@volterra.mat.uniroma2.it
\\
$\ddagger$Department of Information Sciences,
Tokyo University of Science,  Noda City, Chiba 278-8510,
Japan
\\
e-mail: ohya@is.noda.tus.ac.jp
}

\maketitle

\medskip
]

\section{Introduction}

Although the ability of computer is highly progressed, there are several
problems which may not be solved effectively, namely, in polynomial time.
Among such problems, NP problem and NP complete problem are fundamental. It
is known that all NP complete problems are equivalent and an essential
question is{\it \ whether there exists an algorithm to solve an NP complete
problem in polynomial time}. They have been studied for decades and for
which all known algorithms have an exponential running time in the length of
the input so far. The standard definition of P- and NP-problems is
the following \cite{GJ,Cle,[Calud02],OV3}:

\begin{definition}
Let $n$ be the size of input. \newline
(1)A P-problem is a problem whose time needed for solving the problem is
polynomial time of $n.$ Equivalently, it is a problem which can be
recognized in a polynomial time of $n$ by deterministic Turing machine. 
\newline
(2)An NP-problem is a problem that can be solved in polynomial time by a
nondeterministic Turing machine.
\end{definition}

This can be understood as follows: Let consider a problem to find a solution
of $f\left( x\right) =0$. We can check in\ polynomial time of $n$ whether $%
x_{0}$ is a solution of $f\left( x\right) =0$, but we do not\ know whether
we can find the solution of $f\left( x\right) =0$ in polynomial time of $n$.

\begin{definition}
An NP-complete problem is a problem polynomialy transformed NP-problem.
\end{definition}

We take the SAT (satisfiable) problem, one of the NP-complete problems, to
study whether there exists an algorithm showing NPC=P. Our aim in this paper
and the previous papers \cite{OM,OV1,OV2} is to find a quantum algorithm solving
the SAT problem in polynominal time of the size of the problem. 

Let $X\equiv \left\{ {x_{1},\cdots ,x_{n}}\right\} $ be a set. Then $x_{k}$
and its negation $\bar{x}_{k}\left( {k=1,2,\cdots ,n}\right) $ are called
literals and the set of all such literals is denoted by $X^{\prime }\equiv
\left\{ {x_{1},\bar{x}_{1},\cdots ,x_{n},\bar{x}_{n}}\right\} $. The set of
all subsets of $X^{\prime }$ is denoted by ${\cal F}\left( {X^{\prime }}%
\right) $ and an element $C\in {\cal F}\left( {X^{\prime }}\right) $ is
called a clause. We take a truth assignment to all Boolean variables $x_{k}$%
. If we can assign the truth value to at least one element of $C$, then $C$
is called satisfiable. When $C$ is satisfiable, the truth value $t\left(
C\right) $ of $C$ is regarded as true, otherwise, that of $C$ is false. Take
the truth values as ''true $\leftrightarrow $1, false $\leftrightarrow $0''.
Then $C$is satisfiable iff $t\left( C\right) =1$.

Let $L=\left\{ {0,1}\right\} $ be a Boolean lattice with usual join $\vee $
and meet $\wedge $, and $t\left( x\right) $ be the truth value of a literal $%
x$ in $X$. Then the truth value of a clause $C$ is written as $t\left(
C\right) \equiv \vee _{x\in C}t\left( x\right) $.

Moreover the set ${\cal C}$ of all clauses $C_{j}\left( {j=1,2,\cdots ,m}%
\right) $ is called satisfiable iff the meet of all truth values of $C_{j}$
is 1; $t\left( {\cal C}\right) \equiv \wedge _{j=1}^{m}t\left( {C_{j}}%
\right) =1$. Thus the SAT problem is written as follows:

\begin{definition}
SAT Problem: Given a Boolean set $X\equiv \left\{ {x_{1},\cdots ,x_{n}}%
\right\} $and a set ${\cal C}=\left\{ {\cal C}_{1},\cdots ,{\cal C}_{m}
\right\} $ of clauses, determine whether ${\cal C}$ is satisfiable or not.
\end{definition}

That is, this problem is to ask whether there exists a truth assignment to
make ${\cal C}$ satisfiable. It is known in usual algorithm that it is
polynomial time to check the satisfiability only when a specific truth
assignment is given, but we can not determine the satisfiability in
polynomial time when an assignment is not specified.

Ohya and Masuda pointed out \cite{OM} that the SAT problem, hence all other
NP problems, can be solved in polynomial time by quantum computer if the
superposition of two orthogonal vectors $\left\vert 0\right\rangle $ and $%
\left\vert 1\right\rangle $ can physically detected. This result 
was rewritten in \cite{AS} showing that OM SAT-algorithm is combinatoric. 

The output of the OM quantum--SAT algorithm is a superposition vector
$\alpha \left\vert 0\right\rangle +\beta\left\vert 1\right\rangle ,$ 
and, in order to effectively implement this algorithm, it is necessary 
to distinguish this superposition from the pure vector  $\left\vert 0\right\rangle $.
If $\beta $ is not zero but very small this detection is considered not to be 
possible with the present technology.  In \cite{OV2} it is shown that such a 
distinction can be realized by combining a nonlinear chaos amplifier with
the OM quantum algorithm, which implies the existence of a mathematical
algorithm solving NP=P. 
It is not known if the amplification method of Ohya and Volovich is
in the framework of quantum Turing algorithm or not. So the next question
is (1) whether there exists a physical realization combining the quantum--SAT 
algorithm with chaos dynamics, or (2) whether there exists another method to
achieve the above distinction of two vectors by a suitable unitary evolution
so that all processes can be discussed by a quantum Turing machine (circuit).
In this paper, we argue that the stochastic limit \cite{ALV}, can be used to find another
method to solve problem (2) above. 

In Section 2 we review the mathematical frame of the OM quantum--SAT
algorithm and in Section 3 we review the representation of this algorithm 
given by Accardi and Sabaddini \cite{AS}. In Section 4, we state the problem to
distinguish two vectors with a quick review of OV-chaos algorithm. In
Section 5 the new notion of quantum adaptive stochastic system is proposed and
we show that it can be used to solve the problem NPC=P. The details (e.g.,
proofs) of this paper is discussed in \cite{AO2}.

\section{Quantum Algorithm}

The quantum algorithms discussed so far are rather idealized because
computation is represented by unitary operations. A unitary operation is
rather difficult to realize in physical processes, more realistic operation
is one allowing some dissipation like semigroup dynamics. However such
dissipative dynamics destroys the entanglement hence they very much reduce 
the ability of quantum computation of preserving the entanglement of states. 
In order to keep the high ability of quantum computation and good entanglement, 
it will be necessary to introduce some kind of amplification in the course of 
real physical processes in physical devices,
which will be similar to the amplication processes in quantum communication. 
In this section, to search for more realistic operations in
quantum computer, the channel expression will be useful, at least, in the
sense of mathematical scheme of quantum computation because the channel is
not always unitary and represents many different types of dynamics.

Let ${\cal H}$ be a Hilbert space describing input, computation and output
(result). As usual, the Hilbert space is ${\cal H}=\otimes _{1}^{N}{\bf C}%
^{2}$, and let the basis of ${\cal H}=\otimes _{1}^{N}{\bf C}^{2}$ be: $%
e_{0}\left( {=\left| 0\right\rangle }\right) =\left| 0\right\rangle \otimes
\cdots \otimes \left| 0\right\rangle \otimes \left| 0\right\rangle
,e_{1}\left( {=\left| 1\right\rangle }\right) =\left| 0\right\rangle \otimes
\cdots \otimes \left| 0\right\rangle \otimes \left| 1\right\rangle ,\cdots
,e_{2^{N}-1}\left( {=\left| {2^{N}-1}\right\rangle }\right) =\left|
1\right\rangle \otimes \cdots \otimes \left| 1\right\rangle \otimes \left|
1\right\rangle .$

Any number $t$ $\left( {0,\cdots ,2^{N}-1}\right) $ can be expressed by 
$$%
t=\sum\limits_{k=1}^{N}{a_{t}^{\left( k\right) }}2^{k-1},$$ 
${a_{t}^{\left(k\right) }=0}$
${or} \ = 1$, so that the associated vector is written by

\[
\left| t\right\rangle \left( {=e_{t}}\right) =\otimes _{k=1}^{N}\left| {%
a_{t}^{\left( k\right) }}\right\rangle . 
\]
And applying n-tuples of Hadamard matrix $A\equiv \frac{1}{\sqrt{2}}\left( 
\begin{array}{cc}
1 & 1 \\ 
1 & -1%
\end{array}
\right) $ to the vacuum vector $\left| 0\right\rangle ,$ we get 
$$A\left|
0\right\rangle \left( {\ =\xi \left( 0\right) }\right) \equiv \otimes
_{1}^{N}\frac{1}{\sqrt{2}}\left( {\left| 0\right\rangle +\left|
1\right\rangle }\right) .$$

 Put

\[
W\left( t\right) \equiv \otimes _{j=1}^{N}\left( 
\begin{array}{cc}
1 & 0 \\ 
0 & \exp (\frac{2\pi it}{2^{N}}2^{j-1})%
\end{array}
\right) . 
\]
Then we have

\[
\xi \left( t\right) \equiv W\left( t\right) \xi \left( 0\right) =
\]
\[
\frac{1}{%
\sqrt{2^{N}}}\sum\limits_{k=0}^{2^{N}-1}{exp\left( {\frac{{2\pi itk}}{{2^{N}}%
}}\right) }\left| k\right\rangle , 
\]
which is called Discrete Fourier Transformation. The combination of the
above operations gives a unitary operator $U_{F}\left( t\right) \equiv
W\left( t\right) A$ and the vector $\xi \left( t\right) =U_{F}\left(
t\right) \left| 0\right\rangle .$

\subsection{Channel expression of conventional unitary algorithm}

All conventional unitary algorithms can be written as a combination 
of the following three steps:

(1) Preparation of state: Take a state $\rho $ (e.g., $\rho =\left|
0\right\rangle \left\langle 0\right| $) and apply the unitary channel defined
by the above $U_{F}\left( t\right) :\Lambda _{F}^{\ast }=Ad_{U_{F}\left(
t\right) }$%
\[
\Lambda _{F}^{\ast }=Ad_{U_{F}}\Longrightarrow \Lambda _{F}^{\ast }\rho
=U_{F}\rho U_{F}^{\ast } 
\]

(2) Computation: Let $U$ a unitary operator on ${\cal H}$ representing the
computation followed by a suitable programming of a certain problem, then
the computation is described by a channel $
\Lambda _{U}^{\ast }=Ad_{U} $ 
(unitary channel). 
\quad After the computation, the final state $\rho _{f}$ will be

\[
\rho _{f}={\Lambda _{U}^{\ast }\Lambda _{F}^{\ast }\rho .} 
\]

(3) Register and Measurement: For the registration of the computed result and
its measurement we might need an additional system ${\cal K}$ (e.g.,
register), so that the lifting ${\cal E}_{m}^{\ast }$ from ${\cal S}\left( 
{\cal H}\right) $ to ${\cal S}\left( {{\cal H}\otimes {\cal K}}\right) $ in
the sense of \cite{AO} is useful to describe this stage. Thus the whole
process is wrtten as$\ \ \ \ \ $

\[
\rho _{f}={\cal E}_{m}^{\ast }\left( {\Lambda _{U}^{\ast }\Lambda _{F}^{\ast
}\rho }\right) . 
\]%
Finally we measure the state in ${\cal K}$: For instance, let $\left\{ {%
P_{k};k\in J}\right\} $ be a projection valued measure (PVM) on ${\cal K}$%
\[
\Lambda _{m}^{\ast }\rho _{f}=\sum\limits_{k\in J}I\otimes {P_{k}}\rho
_{f}I\otimes P_{k}, 
\]%
after which we can get a desired result by observations in finite times if
the size of the set $J$ is small.

\subsection{Channel expression of the general quantum algorithm}

When dissipation is involved the above three steps have to be generalized. 
Such a generalization can be expressed by means of suitable channel, not
necessarily unitary.

(1) Preparation of state: We may be use the same channel $\Lambda _{F}^{\ast
}=Ad_{U_{F}}$ in this first step, but if the number of qubits $N$ is large
so that it will not be built physically, then $\Lambda _{F}^{\ast }\ $should
be modified, and let denote it by $\Lambda _{P}^{\ast }.$

(2) Computation: This stage is certainly modified to a channel $\Lambda
_{C}^{\ast }$ reflecting the physical device for computer.

(3) Registering and Measurement: This stage will be remained as aobe. Thus
the whole process is written as$\ \ \ \ \ $

\[
\rho _{f}={\cal E}_{m}^{\ast }\left( {\Lambda _{C}^{\ast }\Lambda _{P}^{\ast
}\rho }\right) . 
\]

\section{Quantum Algorithm of SAT}

Let 0 and 1 of the Boolean lattice $L$ be denoted by the vectors $\left|
0\right\rangle \equiv \left( 
\begin{array}{l}
1 \\ 
0%
\end{array}
\right) $ and $\left| 1\right\rangle \equiv \left( 
\begin{array}{l}
0 \\ 
1%
\end{array}
\right) $ in the Hilbert space {\bf C}$^{2},$ respectively. That is, the
vector $\left| 0\right\rangle $ corresponds to falseness and $\left|
1\right\rangle $ does to truth. This section is based on \cite{OM,AS,AO2}.

As we explained in the previous section, an element $x\in X$ can be denoted
by 0 or 1, so by $\left| 0\right\rangle $ or $\left| 1\right\rangle .$ In
order to describe a clause $C$ with at most $n$ length by a quantum state,
we need the n-tuple tensor product Hilbert space ${\cal H\equiv }$ $\otimes
_{1}^{n}${\bf C}$^{2}.$ For instance, in the case of $n=2$, \ given $%
C=\left\{ x_{1},x_{2}\right\} $ with an assignment $x_{1}=0$ and $x_{2}=1,$
then the corresponding quantum state vector is $\left| 0\right\rangle
\otimes \left| 1\right\rangle ,$ so that the quantum state vector describing 
$C$ is generally written by $\left| C\right\rangle =\left|
x_{1}\right\rangle \otimes \left| x_{2}\right\rangle \in $ ${\cal H}$ with $%
x_{k}=0$ or $1$ (k=1,2).

The quantum computation is performed by a unitary gate constructed from
several fundamental gates such as Not gate, Controlled-Not gate,
Controlled-Controlled Not gate\cite{EJ,O1}. Once $X\equiv \left\{
x_{1},\cdots ,x_{n}\right\} $ and ${\cal C=}\left\{ C_{1},C_{2},\cdots
,C_{m}\right\} $ are given, the SAT is to find the vector

\[
\left| t\left( {\cal C}\right) \right\rangle \equiv \wedge _{j=1}^{m}\vee
_{x\in C_{j}}t(x), 
\]
where $t(x)$ is $\left| 0\right\rangle \ $or $\left| 1\right\rangle $ when $%
x=0$ or 1, respectively, and $t(x)\wedge t(y)\equiv t(x\wedge y)$, $t(x)\vee
t(y)\equiv t(x\vee y).$

\subsection{ Logical negation}

\begin{definition}
Let $X$ be a set. A {\it negation\/} on $X$ is an involution without fixed
points, i.e. a map $x\in X\mapsto x^{\prime }\in X$ such that 
$(x^{\prime })^{\prime }=x\quad ;
x\not=x^{\prime } \quad 
\forall \,x\in X $, 
$x^{\prime }$ is called the {\it negation\/} of $x$.
\end{definition}

\begin{proposition}
\label{Pr1.1} Given a nonempty set $X$ with a negation $(x\mapsto x^{\prime
})$ and denoting, for $I\subseteq X$ 
\[
I^{\prime }:=\{x^{\prime }\in X:x\in I\} 
\]
there exists a set $I\subseteq X$ such that $X=I\cup I^{\prime }.$
\end{proposition}

\noindent \quad Thus a finite set with a negation must be even. \label%
{canpart} Let $X$ be a finite set with $2n$ elements and with a negation $%
(x\mapsto x^{\prime })$. A partition 
$X=I\cup I^{\prime }\quad ,\quad |I|=n 
$ can be constructed with an $n$--step algorithm.

\noindent \quad\ Not all $n$--step algorithms are equivalent. The problem
may come from the following question: having produced $k$ elements of a
given set, how difficult is to produce a mean element $x_{k+1}$, different
from the previous ones?

\begin{definition}
\label{mincl} Given a set $X$ with a negation $x\mapsto x^{\prime}$, a
``clause'' is a subset of $X$. A minimal clause is a subset $I\subseteq X$
such that 
$I\cap I^{\prime}=\phi 
$ (i.e. if $I$ contains $x$, it does not contain the negation of $x$).
\end{definition}

In a set $X$ of cardinality $2n$ there are $2^{n}$ minimal clauses. Given a
set $\hat{{\cal C}}_{0}$ of clauses, if there are non minimal clauses in it,
then we can eliminate then from $\hat{{\cal C}}_{0}$ because any truth
function must be identically zero on a non minimal clause.

However, to eliminate the non minimal clauses from $\hat{{\cal C}}_0$, one
has to ``read'' all its elements. These can be of order $2^n$.

\subsection{ Truth functions}

\medskip The set $\{0,1\}$ is a boolean algebra with the operations 
\[
\varepsilon \vee \varepsilon ^{\prime }:=\max \{\varepsilon ,\varepsilon
^{\prime }\},\text{ }\varepsilon \wedge \varepsilon ^{\prime }:=\min
\{\varepsilon ,\varepsilon ^{\prime }\} 
\]
$(\varepsilon ,\varepsilon ^{\prime }\in \{0,1\})$. A clause truth function
on the clauses on the set $\{x_{1},\dots ,x_{n},x_{1}^{\prime },\dots
,x_{n}^{\prime }\}$ is a boolean algebra homomorphism 
\[
t:\hbox{ Parts of }\{x_{1},\dots ,x_{n},x_{1}^{\prime },\dots ,x_{n}^{\prime
}\}\rightarrow \{0,1\} 
\]
with the property ({\it principle of the excluded third \/}): 
$$
t(x_{j})\vee t(x_{j}^{\prime })=1\ ;\quad \forall \,j=1,\dots ,n\eqno(1) 
$$

Because of (1), such a function is uniquely determined by the values $%
\{t(x_{1}),\dots ,t(x_{n})\},$ hence such functions are $2^{n}$. For this
reason, in the following, we will simply say {\it truth function on \/} $%
\{x_{1},\dots ,x_{n}\}$ meaning by this a truth function on the clauses of
the set $\{x_{1},\dots ,x_{n},x_{1}^{\prime },\dots ,x_{n}^{\prime }\}$.
Conversely given any $n$--ple $\varepsilon =(\varepsilon _{1},\dots
,\varepsilon _{n})\in \{0,1\}^{n},$ there exists only one truth function on $%
\{x_{1},\dots ,x_{n}\}$, with the property that 
\[
t(x_{j})=\varepsilon _{j}\ ;\quad \forall \,j=1,\dots ,n 
\]
In the following, given a truth function $t$, we will denote $\varepsilon
_{t}$ the string in $\{t(x_{1}),\dots ,t(x_{n})\}$ uniquely associated to
that function.\bigskip

\quad \noindent Let ${\cal T}$ be the set of truth functions on $%
\{x_{1},\dots ,x_{n}\}$. The function 
\[
t\in {\cal T}\mapsto |t(x_{1}),\dots ,t(x_{n})\rangle \in \otimes ^{n}{\bf %
C^{2}} 
\]
defines a one--to--one correspondence between ${\cal T}$ and the set $%
\{0,1\} $, that is, a one--to--one correspondence between truth functions
and vectors of the computational basis of $\otimes ^{n}{\bf C^{2}}$)

\begin{proposition}
\noindent Let $C\subseteq X$ be a clause and $I$, $I^{\prime }$ the sets
associated to it through the procedure explained in \S\ (1). Let $t$ be a
truth function on $\{x_{1},\dots ,x_{n}\}$. Then 
\[
t(C)=\left[ \bigvee_{i\in I}t(x_{i})\right] \vee \left[ \bigvee_{j\in
I^{\prime }}(1-t(x_{j}))\right] 
\]
\bigskip
\end{proposition}

Therefore as stated in Introduction, \noindent a set of clauses ${\cal C}%
_{0} $ is said to be SAT if there exists a truth function $t$, on $%
\{x_{1},\dots ,x_{n}\}$ such that 
\[
t({\cal C}_{0}):=t(\wedge _{C\in {\cal C}_{0}}C)=\prod_{C\in {\cal C}%
_{0}}t(C)=1 
\]

\subsection{ Quantum algorithm for the SAT Problem}

We review here a technique, developed in \cite{OM}, which shows that the SAT
problem can be solved in polynomial time by a quantum computer.

Given a set of clauses ${\cal C}_{0}=\{C_{1},\dots ,C_{m}\}$ on $X$, Ohya
and Masuda construct a Hilbert space ${\cal H}=\otimes ^{n+\mu }{\bf C^{2}}$
where $\mu $ is a number that can be chosen linear in $mn$, and a unitary
operator $U_{{\cal C}_{0}}:{\cal H}\rightarrow {\cal H}$ with the property
that, for any truth function $t$, 
\[
U_{{\cal C}_{0}}|\varepsilon _{t},0_{\mu }>=|\varepsilon _{t},x_{\mu
-1}^{\varepsilon _{t}},t({\cal C}_{0})> 
\]
where, $\varepsilon _{t}$ is the vector of the computational basis of $%
\otimes ^{n}{\bf C^{2}}$ corresponding to $t$, and $0_{\mu }$ (resp. $x_{\mu
-1}^{\varepsilon }$) is a string of $\mu $ zeros (resp. a string of $(\mu
-1) $ binary symbols depending on $\varepsilon $).

Furthermore $U_{{\cal C}_0}$ is a product of {\it gates\/}, namely of
unitary operators that act at most on two q--bits each time.

Let ${\cal C}_{0}$ and $U_{{\cal C}_{0}}$ be as above and, for every $%
\varepsilon \in \{0,1\}^{n}$, let $t_{\varepsilon }$ be the corresponding
truth function. Applying the unitary operator $U_{{\cal C}_{0}}$ to the
vector 
\[
\left| v\right\rangle :={\frac{1}{2^{n/2}}}\,\sum_{\varepsilon \in
\{0,1\}^{n}}|\varepsilon ,0_{\mu }> 
\]
one obtains the final state vector $|v_{f}>$ 
\[
|v_{f}>:=U_{{\cal C}_{0}}\left| v\right\rangle ={\frac{1}{2^{n/2}}}%
\,\sum_{\varepsilon \in \{0,1\}^{n}}|\varepsilon ,x_{\mu -1}^{\varepsilon
},t_{\varepsilon }({\cal C}_{0})> 
\]

\begin{theorem}
\noindent\ ${\cal C}_{0}$ is SAT if and only if, 
\[
P_{n+\mu ,1}U_{{\cal C}_{0}}\left| v\right\rangle \not=0 
\]
where $P_{n+\mu ,1}$ denotes the projector 
\[
P_{n+\mu ,1}:=1_{n+\mu -1}\otimes |1><1| 
\]
onto the subspace of ${\cal H}$ spanned by the vectors 
\[
|\varepsilon _{n},\varepsilon _{\mu -1},1> 
\]
\end{theorem}

According to the standard theory of quantum measurement, after a measurement
of the event $P_{n+\mu ,1}$, the state 
$\rho =|v_{f}><v_{f}| $ 
becomes 
\[
\rho \rightarrow {\frac{P_{n+\mu ,1}\rho P_{n+\mu ,1}}{Tr\rho ^{\prime
}P_{n+\mu ,1}}}=:\rho ^{\prime } 
\]
Thus the solvability of the SAT problem is reduced to check that $\rho
^{\prime }\neq 0$. The difficulty is that the probability of $P_{n+\mu ,1}$
is 
\[
Tr\rho ^{\prime }P_{n+\mu ,1}=\Vert P_{n+\mu ,1}\psi \Vert ^{2}={\frac{|T(%
{\cal C}_{0})|}{2^{n}}} 
\]
where $|T{\cal C}_{0})|$ is the cardinality of the set $T({\cal C}_{0}) $,
of all the truth functions $t$ such that $t({\cal C}_{0})=1.$

{\em We put }$q:=${\em \ }$\sqrt{{\frac{r}{2^{n}}}}$ {\em with} $r:=|T({\cal %
C}_{0})|$ $in${\em \ the sequel. Then if }$r${\em \ is suitably large to
detect it, then the SAT problem is solved in polynomial time. However, for
small }$r,${\em \ the probability is very small and this means we in fact
don't get an information about the existence of the solution of the equation 
}$t(C_{0})=1,${\em \ so that in such a case we need further deliberation.}

\quad Let us simplify our notations. After the quantum computation, the
quantum computer will be in the state 
\[
\left| v_{f}\right\rangle =\sqrt{1-q^{2}}\left| \varphi _{0}\right\rangle
\otimes \left| 0\right\rangle +q\left| \varphi _{1}\right\rangle \otimes
\left| 1\right\rangle 
\]
where $\left| \varphi _{1}\right\rangle $ and $\left| \varphi
_{0}\right\rangle $ are normalized $n$ qubit states and $q=\sqrt{r/2^{n}}.$
Effectively our problem is reduced to the following $1$ qubit problem. We
have the state 
\[
\left| \psi \right\rangle =\sqrt{1-q^{2}}\left| 0\right\rangle +q\left|
1\right\rangle 
\]
and we want to distinguish between the cases $q=0$ and $q>0$(small positive
number).

\quad It is argued in \cite{BBBV} that quantum computer can speed up {\bf NP}
problems quadratically but not exponentially. The no-go theorem states that
if the inner product of two quantum states is close to 1, then the
probability that a measurement distinguishes which one of the two it is
exponentially small. And one could claim that amplification of this
distinguishability is not possible.

\quad The proposal of \cite{OV2} is that we do not make
a measurement, which will be overwhelmingly likely to fail,
but to use the output $I\left| \psi \right\rangle $ of the quantum
computer as an input for another device which uses classical chaotic 
dynamics to amplify the required probability.

Such a chaos amplifier introduces a new ingredient in quantum computation
and in \cite{OV1,OV2} it is proved  that such an 
amplification is possible in polynomial time.

\quad 
In \cite{OV1,OV2} a practical realization of the new amplifier mechanism 
is not suggested, however it seems to us that the quantum chaos amplifier
considered in \cite{OV2} deserves an investigation and has a potential to be realizable.

\subsection{Chaotic dynamics}

Various aspects of classical and quantum chaos have been the subject of
numerous studies, see \cite{O2} and ref's therein. Here we will argue that
chaos can play a constructive role in computations (see \cite{OV1,OV2} for
the details).

\quad Chaotic behavior in a classical system usually is considered as an
exponential sensitivity to initial conditions. It is this sensitivity we
would like to use to distinguish between the cases $q=0$ and $q>0$ from the
previous section.

\quad Consider the so called logistic map which is given by the equation 
\[
x_{n+1}=ax_{n}(1-x_{n})\equiv f(x),~~~x_{n}\in \left[ 0,1\right] . 
\]

\noindent \noindent \noindent The properties of the map depend on the
parameter $a.$ If we take, for example, $a=3.71,$ then the Lyapunov exponent
is positive, the trajectory is very sensitive to the initial value and one
has the chaotic behavior \cite{O2}. It is important to notice that if the
initial value $x_{0}=0,$ then $x_{n}=0$ for all $n.$

\quad It is known \cite{Deu} that any classical algorithm can be implemented
on quantum computer. Our quantum chaos computer will be consisting from two
blocks. One block is the ordinary quantum computer performing computations
with the output $\left| \psi \right\rangle =\sqrt{1-q^{2}}\left|
0\right\rangle +q\left| 1\right\rangle $. The second block is a computer
performing computations of the {\it classical} logistic map. This two blocks
should be connected in such a way that the state $\left| \psi \right\rangle $
first be transformed into the density matrix of the form 
\[
\rho =q^{2}P_{1}+\left( 1-q^{2}\right) P_{0} 
\]
where $P_{1}$ and $P_{0}$ are projectors to the state vectors $\left|
1\right\rangle $ and $\left| 0\right\rangle .$ This connection is in fact
nontrivial and actually it should be considered as the third block. One has
to notice that $P_{1}$ and $P_{0}$ generate an Abelian algebra which can be
considered as a classical system. In the second block the density matrix $%
\rho $ above is interpreted as the initial data $\rho _{0}$, and we apply
the logistic map as 
\[
\rho _{m}=\frac{(I+f^{m}(\rho _{0})\sigma _{3})}{2} 
\]
where $I$ is the identity matrix and $\sigma _{3}$ is the z-component of
Pauli matrix on ${\bf C}^{2}.$ \ To find a proper value $m$ we finally
measure the value of $\sigma _{3}$ in the state $\rho _{m}$ such that

\[
M_{m}\equiv tr\rho _{m}\sigma _{3}. 
\]
We obtain

\begin{theorem}
\quad 
\[
\rho _{m}=\frac{(I+f^{m}(q^{2})\sigma _{3})}{2},\text{ and }%
M_{m}=f^{m}(q^{2}). 
\]
\end{theorem}

\quad Thus the question is whether we can find such a $m$ in polynomial
steps of $n\ $satisfying the inequality $M_{m}\geq \frac{1}{2}$ for very
small but non-zero $q^{2}.$ Here we have to remark that if one has $q=0$
then $\rho _{0}=P_{0}$ and we obtain $M_{m}=0$ for all $m.$ If $q\neq 0,$
the stochastic dynamics leads to the amplification of the small magnitude $q$
in such a way that it can be detected as is explained below. The transition
from $\rho _{0}$ to $\rho _{m}$ is nonlinear and can be considered as a
classical evolution because our algebra generated by $P_{0}$ and $P_{1}$ is
abelian. The amplification can be done within at most 2n steps due to the
following propositions. Since $f^{m}(q^{2})$ is $x_{m}$ of the logistic map $%
x_{m+1}=f(x_{m})$ with $x_{0}=q^{2},$ we use the notation $x_{m}$ in the
logistic map for simplicity.

\begin{theorem}
For the logistic map $x_{n+1}=ax_{n}\left( 1-x_{n}\right) $ with $a$ $\in %
\left[ 0,4\right] $ and $x_{0}\in \left[ 0,1\right] ,$ let $x_{0}\ $be $%
\frac{1}{2^{n}}$ and a set $J\ $be $\left\{ 0,1,2,\cdots ,n,\cdots
2n\right\} .$ If $a$ is $3.71,$ then there exists an integer $m$ in $J$
satisfying $x_{m}>\frac{1}{2}.$
\end{theorem}

\begin{theorem}
Let $a$ and $n$ be the same in the above proposition. If there exists $m_{0}$
in $J$ such that $x_{m_{0}}>\frac{1}{2}$ $,$ then $m_{0}>\frac{n-1}{\log
_{2}3.71}.$
\end{theorem}

\quad According to these theorems, it is enough to check the value $x_{m}$ $%
(M_{m})$ around the above $m_{0}$ when $q$ is $\frac{1}{2^{n}}$ for a large $%
n$. More generally, when $q$=$\frac{k}{2^{n}}$ with some integer $k,$ it is
similarly checked that the value $x_{m}$ $(M_{m})$ becomes over $\frac{1}{2}$
within at most 2n steps.

The complexity of the quantum algorithm for the SAT problem was discussed in
Section 3 to be in polynomial time. We have only to consider the number of
steps in the classical algorithm for the logistic map performed on quantum
computer. It is the probabilistic part of the construction and one has to
repeat computations several times to be able to distinguish the cases $q=0$
and $q>0.$ Thus it seems that the quantum chaos computer can solve the SAT
problem in polynomial time.

In conclusion of \cite{OV2}, the quantum chaos computer combines the
ordinary quantum computer with quantum chaotic dynamics amplifier. It may go
beyond the usual quantum Turing algorithm, but such a device can be powerful
enough to solve the {\bf NP}-complete problems in the polynomial time. The
detail estimation of the complexity of the SAT algorithm is discussed in 
\cite{IA}.

In the next two sections we will discuss the SAT problem in a different
view, that is, we will show that the same amplification is possible by
unitary dynamics defined in the stochastic limit.

\section{ Quantum Adaptive Systems}

In this section we begin to develop our programmeof constructing a
physically realizable quantum amplifier for the OM quantum--SAT algorithm
which is entirely within the frame of standard quantum computation, namely:
unitary evolutions implemented by the usual physical interactions (in fact we will
consider the simplest class of these interactions: the dipole type ones).

A classical amplifier is a sensor which reacts differently to the state
of the input system, in other words it is an adaptive system. 
In our case the input state is the output of the OM algorithm, which is a 
quantum superposition. Thus we need a quantum amplifier and this naturally 
leads to the problem of developing a physically ealizable notion of quantum
adaptive system.

The idea to develop a mathematical approach to adaptive systems, i.e. those
systems whose properties are in part determined as responses to an
environment \cite{AI,KOT}, were born in connection with some problems of
quantum measurement theory and chaos dynamics.

The mathematical definition of adaptive system is in terms of observables,
namely: {\it an adaptive system is a composite system whose interaction
depends on a fixed observable} (typically in a measurement process, this
observable is the observable one wants to measure). Such systems may be
called {\it observable--adaptive.}

In the present paper we want to extend this point of view by introducing
another natural class of adaptive systems which, in a certain sense, is the
dual to the above defined one, namely the class of {\it state--adaptive}
systems. These are defined as follows: {\it a state--adaptive system is a
composite system whose interaction depends on the state of at least one of
the sub--systems at the instant in which the interaction is switched on}.

Notice that both definitions make sense both for classical and for quantum
systems. Since in this paper we will be interested to an application of
adaptive systems to quantum computation, we will discuss only quantum
adaptive systems, but one should keep in mind that all the considerations
below apply to classical systems as well.

The difference between state--adaptive systems and nonlinear dynamical
systems should be emphasized:

(i) in nonlinear dynamical systems (such as those whose evolution is
described by the Boltzmann equation, or nonlinear Schr\"{o}dinger equation, $%
\dots $, ) the interaction Hamiltonian depends on the state at each time $t$%
: $H_{I}=H_{I}(\rho _{t})$ $;$ $\forall t$ .

(ii) in state--adaptive dynamical systems (such as those considered in the
present paper) the interaction Hamiltonian depends on the state only at each
time $t=0$: $H_{I}=H_{I}(\rho _{0}).$

The latter class of systems describes the following physical situation: at
time $t=-T$ ($T>0$) a system $S$ is prepared in a state $\psi _{-T}$ and in
the time interval $[-T,0]$ it evolves according to a fixed (free) dynamics $%
U_{[-T,0]}$ so that its state at time $0$ is $U_{[-T,0]}\psi _{-T}=:\psi
_{0} $ At time $t=0$ an interaction with another system $R$ is switched on
and this interaction depends on the state $\psi _{0}$: $H_{I}=H_{I}(\psi
_{0}).$

If we interpret the system $R$ as {\it environment}, we can say that the
above interaction describes the response of the environment to the state of
the system $S$.

Now from the general theory of stochastic limit \cite{ALV} one knows that,
under general ergodicity conditions, an interaction with an environment
drives the system to a dynamical (but not necessarily thermodynamical)
equilibrium state which depends on the initial state of the environment and
on the interaction Hamiltonian.

Therefore, if one is able to realize experimentally these state dependent
Hamiltonians, one would be able to drive the system $S$ to a pre--assigned
dynamical equilibrium state depending on the input state $\psi_{0} $.

In the following section we will substantiate the general scheme described
above with an application to the quantum computer approach to the SAT
problem described in the previous sections.

\section{Stochastic Limit and SAT Problem}

We illustrate the general scheme described in the previous section in the
simplest case when the state space of the system is ${\cal H}_{S}\equiv {\bf %
C^{2}}$. We fix an orthonormal basis of ${\cal H}_{S}$ as $\{e_{0},e_{1}\}.$

The unknown state (vector) of the system at time $t=0$ 
\[
\psi :=\sum_{\varepsilon \in \{0,1\}}\alpha _{\varepsilon }e_{\varepsilon
}=\alpha _{0}e_{0}+\alpha _{1}e_{1}\ ;\ \ \Vert \psi \Vert =1. 
\]%
In the case of Sec. 3, $\alpha _{1}$corresponds to $q$ and $e_{j}$ does to $%
\left\vert j\right\rangle $ $\left( j=0,1\right) .$ This vector is taken as
input and defines the interaction Hamiltonian in an external field 
\begin{eqnarray*}
H_{I} &=&\lambda |\psi \rangle \langle \psi |\otimes (A_{g}^{+}+A_{g}) \\
&=&\sum \lambda \alpha _{\varepsilon }\overline{\alpha }_{\varepsilon
}|e_{\varepsilon }\rangle \langle e_{\varepsilon ^{\prime }}|\otimes
(A_{g}^{+}+A_{g})
\end{eqnarray*}%
where $\lambda $ is a small coupling constant. Here and in the following
summation over repeated indices is understood.

The free system Hamiltonian is taken to be diagonal in the $e_{\varepsilon }$%
--basis 
\[
H_{S}:=\sum_{\varepsilon \in \{0,1\}}E_{\varepsilon }|e_{\varepsilon
}\rangle \langle e_{\varepsilon }|=E_{0}|e_{0}\rangle \langle
e_{0}|+E_{1}|e_{1}\rangle \langle e_{1}| 
\]%
and the energy levels are ordered so that $E_{0}<E_{1}.$ Thus there is a
single Bohr frequency $\omega _{0}:=E_{1}-E_{0}>0.$ The $1$--particle field
Hamiltonian is 
\[
S_{t}g(k)=e^{it\omega (k)}g(k) 
\]%
where $\omega (k)$ is a function satisfying the basic analytical assumption
of the stochastic limit. Its second quantization is the free field evolution 
\[
e^{itH_{0}}A^{\pm }ge^{-itH_{0}}=A_{S_{t}g}^{\pm } 
\]%
We can distinguish two cases as below, whose cases correspond to two cases
of Sec. 3, i.e., $q>0$ and $q=0.$

{\bf Case (1)}. If $\alpha _{0},\alpha _{1}\not=0$ , then, according to the
general theory of stochastic limit (i.e., $t\rightarrow t/\lambda ^{2})$ 
\cite{ALV}, the interaction Hamiltonian $H_{I}$ is in the same universality
class as 
\[
\tilde{H}_{I}=D\otimes A_{g}^{+}+D^{+}\otimes A_{g} 
\]%
where $D:=|e_{0}\rangle \langle e_{1}|$ (this means that the two 
interactions have the same stochastic limit). 
The interaction Hamiltonian at time $t$ is then 
\[
\tilde{H}_{I}(t)=e^{-it\omega _{0}}D\otimes A_{S_{t}g}^{+}+\hbox{ h.c.}%
=D\otimes A^{+}(e^{it(\omega (p)-\omega _{0})}g)+\hbox{ h.c.} 
\]%
and the white noise $\left( \left\{ b_{t}\right\} \right) $ Hamiltonian
equation associated, via the stochastic golden rule, to this interaction
Hamiltonian is 
\[
\partial _{t}U_{t}=i(Db_{t}^{+}+D^{+}b_{t})U_{t} 
\]%
Its causally normal ordered form is equivalent to the stochastic
differential equation 
\[
dU_{t}=(iDdB_{t}^{+}+iD^{+}dB_{t}-\gamma _{-}D^{+}Ddt)U_{t}, 
\]%
where $dB_{t}:=b_{t}dt.$

The causally ordered inner Langevin equation is 
\begin{eqnarray*}
dj_{t}(x) &=&dU_{t}^{\ast }xU_{t}+U_{t}^{\ast }xdU_{t}+dU_{t}^{\ast }xdU_{t}
\\
&=&U_{t}^{\ast }(-iD^{+}xdB_{t}-iDxdB_{t}^{+}-\overline{\gamma }%
_{-}D^{+}Dxdt+ixDdB_{t}^{+} \\
&&+ixD^{+}dB_{t}-\gamma _{-}xD^{+}Ddt+\gamma _{-}D^{+}xDdt)U_{t} \\
&=&ij_{t}([x,D^{+}])dB_{t}+ij_{t}([x,D])dB_{t}^{+} \\
&&-(\hbox{Re }\gamma _{-})j_{t}(\{D^{+}D,x\})dt+i(Im\gamma
_{-})j_{t}([D^{+}D,x])dt \\
&&+j_{t}(D^{+}xD)(\hbox{Re }\gamma _{-})dt,
\end{eqnarray*}
where $j_{t}(x):=$ $U_{t}^{\ast }xU_{t}.$ Therefore the master equation is 
\begin{eqnarray*}
{\frac{d}{dt}}\,P^{t}(x) &=&(Im\gamma )i[D^{+}D,P^{t}(x)]-(\hbox{Re}\gamma
_{-})\{D^{+}D,P^{t}(x)\} \\
&&+(\hbox{Re }\gamma _{-})D^{+}P^{t}(x)D
\end{eqnarray*}
where $D^{+}D=|e_{1}\rangle \langle e_{1}|$ and $D^{+}xD=\langle
e_{0},xe_{0}\rangle |e_{1}\rangle \langle e_{1}|.$

The dual Markovian evolution $P_{\ast }^{t}$ acts on density matrices and
its generator is 
\[
L_{\ast }\rho =(Im\gamma _{-})i[\rho ,D^{+}D]-(\hbox{Re }\gamma _{-})\{\rho
,D^{+}D\}+(\hbox{Re }\gamma _{-})D\rho D^{+} 
\]
Thus, if $\rho _{0}=|e_{0}\rangle \langle e_{0}|$ one has 
\[
L_{\ast }\rho _{0}=0 
\]
so $\rho _{0}$ is an invariant measure. From the
Fagnola--Rebolledo criteria (cf. \cite{[QP--PQ XIV]}), it is the unique invariant measure
and the semigroup $\exp (tL_{\ast })$ converges exponentially to it.

{\bf Case (2)}. If $\alpha _{1}=0,$ then the interaction Hamiltonian $H_{I}$
is 
\[
H_{I}=\lambda |e_{0}\rangle \langle e_{0}|\otimes (A_{g}^{+}+A_{g}) 
\]%
and, according to the general theory of stochastic limit, the reduced
evolution has no damping and corresponds to the pure Hamiltonian 
\[
H_{S}+|e_{0}\rangle \langle e_{0}|=(E_{0}+1)|e_{0}\rangle \langle
e_{0}|+E_{1}|e_{1}\rangle \langle e_{1}| 
\]%
therefore, if we choose the eigenvalues $E_{1},E_{0}$ to be integers (in
appropriate units), then the evolution will be periodic.

Since the eigenvalues $E_{1},E_{0}$ can be chosen a priori, by fixing the
system Hamiltonian $H_{S}$, it follows that the period of the evolution can
be known a priori. This gives a simple criterium for the solvability of the
SAT problem because, by waiting a sufficiently long time one can
experimentally detect the difference between a damping and an oscillating
behavior.

A precise estimate of this time can be achieved either by theoretical
methods or by computer simulation. Both methods will be analyzed in the full
paper \cite{AO2}.

\section{Conclusion}

We showed in \cite{OM,OV1,OV2} that we could find an algorithm solving the
SAT problems in polynomial steps by combining a quantum algorithm with a
chaos dynamics. We used the logistic map there, however it is possible to
use other chaotic maps if they can amplify one of two cofficients. In this
short paper we pointed out that it is possible to distinguish two different
states, $\sqrt{1-q^{2}}\left\vert 0\right\rangle +q\left\vert 1\right\rangle 
$ $\left( q\neq 0\right) $ and $\left\vert 0\right\rangle $ by means of the
adaptive dynamics and the stochastic limit. Finally we remark that our
algorithm can be described by deterministic general quantum Turing machine 
\cite{IO,AOV}, whose result is based on the general quantum algorithm
mentioned in Sec.2.

{\bf Acknowledgment} \newline
The authors thank SCAT for financial support of our joint work. 


\begin{references}
\bibitem{ALV} L.Accardi, Y.G. Lu, I. Volovich: Quantum Theory and its
Stochastic Limit. Springer Verlag 2002; Japanese translation,
Tokyo--Springer 2003.

\bibitem{AO} L.Accardi and M.Ohya, Compound channels, transition
expectations, and liftings, Appl. Math. Optim., Vol.39, 33-59, 1999.

\bibitem{AO2} L.Accardi and M.Ohya, A stochastic limit approach to the SAT
problem, in preparation.

\bibitem{AS1} Luigi Accardi, Ruben Sabbadini: On the Ohya--Masuda quantum
SAT Algorithm, in: Proceedings Intern.Conf. ''Unconventional Models of
Computations'', I. Antoniou, C.S. Calude, M. Dinneen (eds.) Springer 2001 ;
Preprint Volterra, N. 432, 2000

\bibitem{AS2} Luigi Accardi, Ruben Sabbadini: A Generalization of Grover's
Algorithm, Proceedings Intern.Conf.: Quantum Information III, Meijo
University, Nagoya, 27--31 mARCH, 2001; World Scientific 2002; qu-phys
0012143; Preprint Volterra, N. 444, 2001

\bibitem{AI} L. Accardi and K. Imafuku: Control of Quantum States by
Decoherence, to appear in Open Systems and Information Dynamics, 2003

\bibitem{OM} M. Ohya and N. Masuda, {\it NP problem in Quantum Algorithm,}
Open Systems and Information Dynamics, Vol.7, No.1, 33-39, 2000.

\bibitem{O1} M. Ohya, {\it Mathematical Foundation of Quantum Computer\/},
Maruzen Publ. Company, 1998.

\bibitem{OV1} M.Ohya and I.V.Volovich, Quantum computing, NP-complete
problems and chaotic dynamics, in: {\em Quantum Information II}, eds. T.Hida
and K.Saito, World Sci. 2000; quant-ph/9912100 and J.Opt.B, 5,No.6 639-642,
2003

\bibitem{OV2} M. Ohya and I.V. Volovich: New quantum algorithm for studying
NP-complete problems, Rep.Math.Phys.,52, No.1,25-33, 2003

\bibitem{GJ} M. Garey and D. Johnson, {\it Computers and Intractability - a
guide to the theory of NP-completeness}, Freeman, 1979.

\bibitem{Sho} P.W. Shor, {\it Algorithm for quantum computation : Discrete
logarithm and factoring algorithm}, Proceedings of the 35th Annual IEEE
Symposium on Foundation of Computer Science, pp.124-134, 1994.

\bibitem{BBBV} C. H. Bennett, E. Bernstein, G. Brassard, U. Vazirani, {\it %
Strengths and Weaknesses of Quantum Computing}, quant-ph/9701001.

\bibitem{[Calud02]}
Cristian S. Calude:
Information and randomness,
Springer (2002) (2d edition)


\bibitem{Cle} R. Cleve, {\it An Introduction to Quantum Complexity Theory},
quant-ph/9906111.

\bibitem{AS} L.Accardi, R.Sabbadini: On the Ohya--Masuda quantum SAT
Algorithm, in: Proceedings International Conference ''Unconventional Models
of Computations'', I. Antoniou, C.S. Calude, M. Dinneen (eds.) Springer 2001

\bibitem{O2} M. Ohya, {\it Complexities and Their Applications to
Characterization of Chaos,} Int. Journ. of Theort. Phy., 37, 495, 1998.

\bibitem{OV3} M.Ohya and I.V. Volovich, {\it Quantum information,
computation, cryptography and teleportation, }Springer (to appear).{\it \ }

\bibitem{Deu} D. Deutsch, {\it Quantum theory, the Church-Turing principle
and the universal quantum computer,} Proc. of Royal Society of London series
A, {\bf 400}, pp.97-117, 1985.

\bibitem{EJ} A. Ekert and R. Jozsa, {\it Quantum computation and Shor's
factoring algorithm,} Reviews of Modern Physics, {\bf 68} No.3,pp.733-753,
1996.

\bibitem{IA} S. Iriyama and S. Akashi,{\it \ Complexity of
Ohya-Masuda-Volovich al;gorithm}, to appear

\bibitem{IO} S. Iriyama and M. Ohya, {\it On generalized Turing machine},
TUS(Tokyo University of Science) preprint, 2003.

\bibitem{AOV} L. Accardi, M. Ohya and I.Volovich, in preparation.

\bibitem{KOT} A.Kossakowski, M.Ohya and Y.Togawa, How can we observe and
describe chaos?, Open System and Information Dynamics 10(3): 221-233, 2003

\bibitem{[QP--PQ XIV]}
Quantum Probability and White Noise Analysis, 
Vol. XIV
Quantum interacting particle systems,
Luigi Accardi, Franco Fagnola (eds.)
World Scientific (2002),
\end{references}
\end{document}